\documentclass[reqno]{amsart}
\usepackage{epsf}
\usepackage{color}
\def\be{\begin{equation}}
\def\ee{\end{equation}}
\def\ba{\begin{array}}
\def\ea{\end{array}}
\def\bea{\begin{eqnarray}}
\def\eea{\end{eqnarray}}
\def\drm{{\mathrm d}}

\newcommand{\nc}{\newcommand}
\nc{\tcb}{\textcolor{blue}}
\nc{\tcr}{}
\nc{\tcg}{\textcolor{green}}

\nc{\qq}{\qquad\qquad}
\nc{\dis}{\displaystyle}
\nc{\ug}{\; = \;}
\nc{\Ebf}{\mbox{\boldmath $E$}}
\nc{\Bbf}{\mbox{\boldmath $B$}}
\nc{\abf}{\mbox{\boldmath $a$}}
\nc{\vbf}{\mbox{\boldmath $v$}}
\nc{\Fbf}{\mbox{\boldmath $F$}}
\nc{\rbf}{\mbox{\boldmath $r$}}
\nc{\Jbf}{\mbox{\boldmath $J$}}
\nc{\rd}{{\rm d}}
\nc{\dtau}{{\rd\tau}}
\nc{\dt}{{\rd t}}
\nc{\omf}{{\frac{\omega}{\omz}}}
\nc{\tz}{\tau_0}

\begin{document}

\vspace{1truecm}

\title{Majorana, Pauling and the quantum theory of the chemical bond}

\author{S. Esposito}
\address{{\it S. Esposito}: I.N.F.N. Sezione di
Napoli, Complesso Universitario di M. S. Angelo, Via Cinthia,
80126 Naples, Italy ({\rm Salvatore.Esposito@na.infn.it})}%

\author{A. Naddeo}
\address{{\it A. Naddeo}: Dipartimento di Fisica ``E.R. Caianiello'',
Universit\`a di Salerno, and CNISM Unit\`a di Ricerca di Salerno, Via Ponte don Melillo, 84084 Fisciano (Salerno), Italy
 ({\rm naddeo@sa.infn.it})}%

\begin{abstract}
We discuss in detail very little known results obtained by Majorana as early as 1931, regarding the quantum theory of the chemical bond in homopolar molecules, based on the key concept of exchange interaction.  After a brief historical overview of the quantum homopolar valence theory, we address the intriguing issues of the formation of the helium molecular ion, He$_{2}^{+}$, and of the accurate description of the hydrogen molecule, H$_{2}$. For the first case, the group theory-inspired approach used by Majorana is contrasted with that more known followed by Pauling (and published few months after that of Majorana), while for the second case we focus on his proposal concerning the possible existence of ionic structures in  homopolar compounds, just as in the hydrogen molecule. The novelty and relevance of Majorana's results in the modern research on molecular and chemical physics is emphasized as well.
\end{abstract}

\maketitle



\section{Introduction}

\noindent The successful description of atomic systems, offered by quantum mechanics in the second half of 1920s, resulted quite soon in the belief that ``the underlying physical laws necessary for the mathematical
theory of a large part of physics and the whole of chemistry [were] completely known'', and -- always according to P.A.M. Dirac -- that ``the difficulty [was] only that the exact application of these laws leads to equations much too complicated to be soluble'' \cite{Dirac}. However, it was realized very soon as well that a quantum description of molecules was {\it not} just a simple extension of atomic physics, whose main problem being only mathematical in nature, and novel physical ideas should necessarily complement appropriate mathematical methods. Quite characteristically, and contrary to what happened for the quantum theory of atoms, the first steps towards a quantum description of molecules were guided by a number of authors contributing with more or less large results, but all of them producing key ingredients for the understanding of the chemical bond. 

The idea of {\it valence} was very early introduced into chemistry\footnote{The concept of a {\it chemical bond} made its first appearance in the chemical literature in 1866, in a paper by E. Frankland \cite{Frank}.} to explain some number relationships in the combining ratios of atoms and ions, but the first attempt to incorporate the atomic structure information in a consistent -- though qualitative -- theoretical framework was performed by G.N. Lewis only in 1916 \cite{Lewis}. In his own view, the chemical bond consisted of a pair of electrons held jointly between two atoms; as summarized by J.H. van Vleck in 1935, 
\begin{quote}
Lewis' theory, based primarily on two ideas -- the idea that each nucleus tends to become surrounded by a closed shell of electrons corresponding to that present in an inert gas atom, and the idea that a pair of electrons shared between two nuclei constitutes the homopolar bond -- made an instant appeal to chemists because it was able to correlate and predict in a simple fashion an enormous number of previously unrelated facts \cite{vleck35}.
\end{quote}
However, Lewis theory didn't say anything on the nature of the forces involved in the formation of the homopolar bond, and only with the advent of quantum mechanics appropriate and powerful theoretical tools became available in order to tackle the problem \cite{Mehra}. In principle, as recalled by Dirac (see above), it became possible in principle to write down an equation for any system of nuclei and electrons, whose solution would provide thorough predictions on the stability of the system under study, but the $n$-body problem revealed to be not amenable to exact analytical solutions, thus triggering the development of several approximation methods. In this scenario the key idea of {\it exchange forces} (or quantum {\it resonance}), introduced in molecular physics by W. Heitler and F. London (and borrowed by W. Heisenberg's theory of  the helium atom \cite{Hei26}), grew up, as opposed to the polarization forces able to describe ionic compounds, and became the basis of the quantum theory of homopolar chemical bond. Indeed, Heitler and London succeeded to account for the stability of the hydrogen molecule and predicted, with remarkable accuracy, the dependence of the total electronic energy on the internuclear distance.

Heitler and London's approach originated from the concept of chemical valence, and was aimed to explain the chemical bond by using properties of the constituent atoms, but since such atoms were assumed (see below) to retain their properties in the molecule to a large extent, their method resulted difficult to apply to molecules more complex than the simple H$_2$. Again, then, by giving a look to what already explored in atomic physics since the times of the old quantum theory, several scientists -- headed mainly by F. Hund, R.S. Mulliken, G. Herzberg and  J.E. Lennard-Jones -- introduced molecular spectroscopy as a guiding principle in developing a theory of the chemical bond. For spectroscopists, used to observe similar properties in many different molecules (formed with different atoms) with the same number of electrons, it was quite natural to assume a molecule to be a collection of nuclei fixed in given spatial positions with an additional electron cloud surrounding them. {\it Molecular orbitals} occupied by each electron were introduced, whose extent stretched over the whole of the molecule, so that the chemical bond resulted from ``sharing'' electrons among the constituent atoms, these then loosing their identity to a large extent, as opposed to the Heitler-London method.

A number of different refinements and generalizations of both approaches appeared in the subsequent literature, but quantitative calculations remained ``much too complicated'' to allow tests of the novel ideas in molecules other than those formed from hydrogen and helium atoms, such calculations being performed mainly within the Heitler-London approach. As pointed out by Herzberg,
\begin{quote}
While the Heitler-London method was developed from a treatment of the H$_2$ molecule, the molecular orbital method may be most readily developed from a study of the H$_2^+$ molecule. The systems H$_2$ and H$_2^+$ are the only ones which have up to the present time been worked out with adequate accuracy on the basis of wave mechanics \cite{Herzberg}.
\end{quote}
While such a conclusion is largely acceptable, however it is often underestimated the values of the contributions developed for molecules composed of {\it helium} atoms. Indeed, although the situation in molecular physics was not completely similar to that in atomic physics, within which the study of the helium atom brought to the definitive success (with respect to the old quantum theory) of quantum mechanics,\footnote{For a review, see Ref. \cite{twoel}.} nevertheless the quantitative study of helium compounds led to relevant results that allowed to test significantly the theory of the chemical bond and to explore their potentialities. Already Heitler and London concluded that no stable He$_2$ could be formed, but the more intricate problem with the helium molecular ion was discussed (and solved) by E. Majorana and, independently, L. Pauling around 1931. What obtained by Pauling, reviewed here in Sect. 3 after a short discussion of (only) the very basic first steps towards the understanding of the quantum nature of the chemical bond, was a direct consequence of the concepts introduced by him of one-electron and three-electron bonds applied to molecules different from the paradigmatic H$_2$ example. The description of the helium molecular ion given by Majorana, and discussed in Sect. 4, was instead a ``true'' quantum mechanical theory based on the relevant symmetry properties of the system \cite{Pucci}, managed by proper group-theory inspired methods, so that -- in a sense -- it enabled to test directly the goodness of the general physical ideas behind the Heitler-London approach without too much overtones.

Also, again in 1931, Majorana contributed with another relevant paper on the chemical bond, by introducing what are now known as ``ionic structures'' in homopolar molecules, this being considered here in Sect. 5.

Such contributions by Majorana do not exhaust the fascinating story of the theory of the chemical bond, but certainly add to it important pieces that are not at all widely known, while their importance to present day research has been disclosed only in recent times (see, for example, the {\it Majorana structures} considered in Ref. \cite{Corongiu}). The present paper is, then, aimed to fill such a gap and to reveal the pregnancy of what already obtained by Majorana for modern molecular physics.


\section{Understanding the quantum nature of the chemical bond}

\noindent A complete history of the first steps towards a quantum description of molecules is beyond the scope of the present paper (and will be published elsewhere), but here we will focus just on few points relevant for our discussion.

The problem with a quantum theory of molecules was quickly recognized to be twofold. On one hand, the differential Schr\"odinger equation applied to a given molecular system does not admit simple analytic solutions even for very simple molecules or ions (such as H$_2^+$, H$_2$, He$_2^+$. etc.), mainly due to the non-central potential acting on the system and to the inter-electronic interactions. Instead, on the other hand, the nature of the chemical bond was not at all clear for homopolar compounds, while some sort of  ``polarization forces'' were assumed for ionic compounds since the times of the old quantum theory.

\subsection{Two-center problems and the Born-Oppenheimer approximation}

The application of standard mathematical techniques customarily exploited in a\-to\-mic physics resulted into even deeper mathematical problems, though some physical insight was at last certainly gained. Two illuminating examples, for instance, referring to the most simple system with one electron in the field of a two-center potential, that is the H$_2^+$ molecular ion, are the following ones. In 1927 Burrau \cite{Burrau} succeeded in numerical integrating the relevant two-center Schr\"odinger equation, within a (divergent) asymptotic expansion, giving explicit numerical predictions for intermediate electronic separations (the electron is neither assumed to be infinitely away from the two nuclei, nor very close to them), but no justification for the use of the asymptotic series was provided. Conversely, Uns\"old \cite{Unsold} adopted a perturbative method, by considering the H$_2^+$ system just as a hydrogen atom ``perturbed'' by the presence of an additional proton, but severe convergence problems were discovered while successive approximations came into play: the second order approximation for the ground state energy yielded a value about twice of that obtained at first order, and with an opposite sign. 

Physical insight into the complex mathematical problem produced solid results by means of the seminal paper by Born and Oppenheimer \cite{BO}. By noticing that the mass $M$ of nuclei in molecules is much larger than that ($m$) of electrons, they ``ordered'' the molecular spectra by realizing that the major contribution to them comes from the electronic motion around the nuclei, followed by nuclear vibrations and, finally, by nuclear rotations around the molecular axis. According to this ``ordering'', they also realized that a thorough approximation of the molecular problem is obtained by expanding the corresponding Hamiltonian $H$ in successive powers of the dimensionless parameter $k=\sqrt[4]{m/M}$, considering (at least) terms of order 4. The approximate ground state energy they wrote (at this order) was then
\be
W = V^{(0)} + k^2 W^{(2)} + k^4 W^{(4)} \, ,
\ee
where $V^{(0)}$ is the minimum value (that is, evaluated at the equilibrium internuclear distance) of the electronic energy, $W^{(2)}$ is the nuclear vibration energy and $W^{(4)}$ is the energy associated to nuclear rotations. The intriguing result was that the three kind of motions are {\it separated} at the order considered, while couplings between them arise only at order greater than fourth (not considered by Born and Oppenheimer). These authors then provided a powerful method to attack molecular problems, which was well described in a review published by Pauling one year later for a (quantum) chemistry audience:
\begin{quote}
The procedure to be followed in the theoretical discussion of the structure of molecules has been given by Born and Oppenheimer, who applied the perturbation theory (to the fourth order) to a system of nuclei and electrons. They showed that the electronic energy is first to be calculated for various arrangements of the nuclei fixed in space. The stable state will then be that for which the so-calculated electronic energy added to the internuclear energy is a minimum. The nuclei will then undergo oscillations about their equilibrium positions, with the electronic and nuclear energy as the restoring potential; and the molecule as a whole will undergo rotations about axes passing through its center of mass \cite{Pauling1928}.
\end{quote}

\subsection{The Heitler-London method}

The physical realization of the chemical bond in homopolar compounds was, instead, firstly recognized by Heitler and London \cite{HL} in the same year 1927. They studied the ``perturbative'' approximation of the Schr\"odinger equation for the hydrogen molecule H$_2$, whose solutions represent neutral H atoms in their ground state (they considered the two nuclei at a fixed distance $R$). The ``unperturbed'' eigenfunctions were chosen in such a way that one of the two $(1,2)$ electrons be on one of the two $(a,b)$ nuclei, and the other electron on the other nucleus; that is, by indicating with $\psi_i$, $\varphi_i$ the eigenfunctions of the H atom, they correspond to the products $\psi_1 \varphi_2$ and $\psi_2 \varphi_1$. Since a twofold degeneracy is present for the hydrogen molecule (two possible ways of assigning the electrons to nuclei), the true unperturbed eigenfunctions are a combination of such products, which, by employing standard orthonormality conditions, Heitler and London wrote as 
\be
\ba{rcl}
& & \dis \alpha = \dis \frac{1}{\sqrt{2+2S}} \left( \psi_1 \varphi_2 + \psi_2 \varphi_1 \right) , \\ & & \\
& & \dis \beta = \dis \frac{1}{\sqrt{2-2S}} \left( \psi_1 \varphi_2 - \psi_2 \varphi_1 \right) 
\ea
\ee
($S = \int \! \psi_1 \varphi_1 \psi_2 \varphi_2 \, \drm \tau_1 \drm \tau_2$ is a normalization integral that will be explicitly calculated later by Sugiura \cite{Sugiura}). They noticed that the eigenfunctions of the atom $a$ does not vanish in the position of the atom $b$ and viceversa, so that a finite probability exist for the electron of nucleus $a$ to belong to the nucleus $b$. In other words, a {\it resonance} phenomenon, similar to that already introduced by Heisenberg for the helium atom \cite{Hei26}, takes place:
\begin{quote}
While in classical mechanics it is possible to label the electrons (we put each electron in a sufficiently steep potential well and do not allow energy addition), something similar is impossible in quantum mechanics: when at one moment in time one is certain to know one electron in the potential well, one can never be certain that in the next moment it does not exchange with another \cite{HL}.
\end{quote}
According to Heitler and London, the ``exchange interaction'' is, thus, responsible for the chemical bond in molecules. Following Heisenberg, they also noticed that the antisymmetric solution corresponding to the energy eigenvalue (as a function of $R$) $E_\beta$ describes a van der Waals repulsion between the two H atoms (the potential energy is ever decreasing with $R$, with no minimum), while these atoms may combine to form a stable molecule if the symmetric solution is excited, the minimum of $E_\alpha$ corresponding to the equilibrium configuration of the system. The authors concluded that ``the non-polar attraction is a characteristic quantum mechanical effect'' \cite{HL}.

They then passed to consider, within the same reasoning, the helium molecule He$_2$, the ``zeroth order'' combinations being the following (by adopting the same notations as above for the four electrons and the two nuclei):
\be
\ba{rcl}
& & \dis \alpha = \psi_{12} \varphi_{34} + \psi_{34} \varphi_{12} + \psi_{14} \varphi_{23} + \psi_{23} \varphi_{14} \, , \\ & & \\
& & \dis \beta = \psi_{12} \varphi_{34} + \psi_{34} \varphi_{12} - \psi_{14} \varphi_{23} - \psi_{23} \varphi_{14} \, , \\ & & \\
& & \dis \gamma = \psi_{12} \varphi_{34} - \psi_{34} \varphi_{12} \, , \\ & & \\ 
& & \dis \delta = \hphantom{\psi_{12} \varphi_{34} + \psi_{34} \varphi_{12} + \, }
\psi_{14} \varphi_{23} - \psi_{23} \varphi_{14} \, .
\ea
\ee
The first two states were recognized to be non degenerate ($E_\alpha \neq E_\beta$), while the opposite is true for the remaining two ($E_\gamma = E_\delta$), which represent not allowed configurations. As for the H$_2$ case, the lowest energy eigenvalue is $E_\alpha$, so that the corresponding state would represent (at the zeroth order) the formation of the helium molecule, while $\beta$ represents the elastic repulsion between the two component atoms. However, Heitler and London showed that the $\alpha$ state did not respect the Pauli principle (two helium atoms, as any two other noble gas atoms, cannot be distinguished according to their spin, as instead happened for the H atoms), so that only the unstable $\beta$ configuration is allowed. A viable solution to this drawback would seem to consider a suitable combination of the four states $\alpha, \beta, \gamma, \delta$, but this is as well not possible, since all the corresponding four electrons would belong to a $K$ shell, which is again prohibited by the Pauli principle: the only solution is a combination with {\it excited} helium atoms. The thorough application of the basic principle of quantum mechanics, following Heisenberg, thus led Heitler and London to the conclusion that no stable He$_2$ molecule can be formed. According to Pauling,
\begin{quote}
It is of particular significance that the straightforward application of the quantum mechanics results in the unambiguous conclusion that two hydrogen atoms will form a molecule but that two helium atoms will not; for this distinction is characteristically chemical, and its clarification marks the genesis of the science of sub-atomic theoretical chemistry \cite{Pauling1928}.
\end{quote}
The physical meaning of what realized by Heitler and London was well understood by Pauling, who ``translated'' it to chemists, and gave them also a visual representation of the chemical bond in the simple H$_2^+$ molecular ion:
\begin{quote}
The electron is most of the time in the region between the two nuclei, and can be considered as belonging to them both, and forming a bond between them \cite{Pauling1928}.
\end{quote}
However, the most profound implications of the Heitler and London theory, further elaborated by the same London \cite{London1928}, who related the mentioned findings to the symmetry properties of the molecular system at hand (and then to group-theoretical properties), were pointed out in a very lucid review of 1929 by van Vleck \cite{Vleck}. He provided a general overview of the molecular problem, not limited to very simple molecules, thus establishing the foundations for the description of any molecular compound:
\begin{quote}
The London and Heitler theory of valences is primarily based on symmetry properties of the wavefunctions and does not aim to say anything about the stability or heats of reaction of the various compounds, as this would require a detailed dynamical investigation. [...] 
\\
The general trend of the work seems to be that because of the critical examination of symmetry properties required by the Pauli exclusion principle, the theory of the classification of valences in complicated organic compounds, etc., must be closely related to the group theory of the mathematicians. [...]
\\
According to London, the reason certain valences or bonds do not occur (e.g., compounds involving inert gases) is not that such bonds lead to molecules which are energetically unstable, but that the bonds, when stable, correspond to solutions of the Schr\"odinger wave equation which are of a type of symmetry contrary to Pauli's exclusion principle. [...]
\\
The different valences correspond to different apportionments of various values of the quantum numbers $k, m_k$ among the electrons, and the relative prevalence of the different valences depends upon the relative prevalence of the states corresponding to different values of the quantum numbers $k, m_k$ but given $n$. Some of these states may have such high energies that they are occupied only very infrequently, and so the corresponding valences may not exist. [...]
\\
London's work, in fact, seems to show that there is a very intimate connection between valences and the spectroscopists' classification of spectral terms \cite{Vleck}.
\end{quote}
The gap between chemistry and spectroscopy was, then, filled -- at least on conceptual grounds --, but a number of particular problems waited for solutions in the subsequent years.

\subsection{Molecular orbitals}

Although the Heitler and London theory represented a watershed in the quantum understanding of the molecular structure, the lacking of analytic computations in molecular problems favored the development of different approximation methods, suitable especially for more complex systems. As noted in 1929 by Lennard-Jones, ``the new theory gives a quantum mechanical explanation of the empirical ideas of Lewis and others concerning chemical valency and correlates the idea of valency with that of electron spin'', but ``certain difficulties have appeared in more complicated cases, as, for instance, in O$_2$ and F$_2$, where the theory is unable to account for molecular binding, and at the same time to account for the $^3\Sigma$ ground level of O$_2$'' \cite{Lennard}. It was just this author, following early ideas from Hund, Mulliken and Herzberg,\footnote{For an appropriate list of references, see what reported by van Vleck in \cite{vleck35}.} that developed the competing method of {\it molecular orbitals}. Here the basic idea was to build the electronic structure of the diatomic molecules analogously to what done for atoms, by starting from the possible energetic levels of a single electron in the presence of a system with two nuclei, and then adding the effect of the other electrons. The key concept of  ``molecular orbital'' played in this novel method, along with a clear analysis of the validity of the method itself, compared with that of Heitler and London, was beautifully explained by van Vleck in his exhaustive review of 1935:
\begin{quote}
A molecular orbital is defined as a wavefunction which is a function of the coordinates of only one electron, and which is, at least hypothetically, a solution of a dynamical problem involving only one electron. The method of
molecular orbitals seeks to approximate the wavefunction of a molecule containing $n$ electrons as the product of $n$ molecular orbitals, so that
\be
\Psi = \psi_1(x_1,y_1,z_1) \psi_2(x_2,y_2,z_2)  \cdots \psi_n(x_n,y_n,z_n) \, .
\ee
[...] The great failing of the method of molecular orbitals is the excessive presence of ionic terms, due to inadequate allowance for the $r_{12}$ repulsion. [...] To avoid this difficulty of inadequate recognition of the $r_{12}$ effect, the Heitler-London method goes to the other extreme, and assumes as its defining characteristic that all ionic terms are completely wanting. [...] The Heitler-London method is much preferable at very large distances of separation of the atoms, at least in symmetrical molecules, for then the continual transfer of electronic charge from one atom to another demanded by the ionic terms surely scarcely occurs at all. On the other hand, at small distances, the Heitler-London method probably represents excessive fear of the $r_{12}$ effect \cite{vleck35}.
\end{quote}
Nevertheless, as correctly pointed out by the same author, the conclusion cannot be that the following:
\begin{quote}
The molecular orbitals are often the more convenient for purposes of qualitative discussion, whereas the Heitler-London method has been used the more frequently for purposes of quantitative calculation \cite{vleck35}.
\end{quote}

\section{One-electron and three-electron bonds}

\noindent Quantitative calculations were performed, as mentioned, for simple diatomic mo\-le\-cu\-les, like H$_2^+$ and H$_2$, but the next step was evidently to understand how the chemical (homopolar) bond can be realized when more electrons are present, according to the requirements imposed by the Pauli exclusion principle. Indeed, the exchange interaction introduced by Heitler and London -- that is, the ``resonance'' phenomenon -- applied to pairs of electrons (as in the H$_2$ molecule) due to their identity, but it could not be applied {\it tout court} to molecular compounds with one or three electrons. Nevertheless, already in 1928 Pauling recognized (see the quotation above) how a one-electron bond could be physically realized in H$_2^+$, and he himself got back to the problem in 1931 \cite{Pauling1931} by explaining that a resonance phenomenon applies also to this molecular ion, since the unperturbed system is degenerate: the two nuclei have the same charge, and thus also the same energy. More in general, he proposed the existence of a one-electron bond according to the following rule:
\begin{quote}
A stable one-electron bond can be formed only when there are two conceivable electronic states of the system with essentially the same energy, the states differing in that for one there is an unpaired electron attached to one atom, and for the other the same unpaired electron is attached to the second atom \cite{Pauling1931}.
\end{quote}
By reasoning on the values of the dissociation energy of several compounds, Pauling suggested also that, in addition to H$_2^+$, one-electron bonds were present in H$_3^+$, Li$_2^+$, boron hydrides, etc., thus opening the road to the understanding of the nature of the chemical bond in more complex systems.

The problem remained for molecular compounds with three electrons,\footnote{As pointed out by Pauling following the Heitler and London reasoning \cite{Pauling1931}, with four electrons there is no tendency to form a strong molecular bond, since two of them are necessarily nuclear symmetric, while the remaining two are nuclear antisymmetric.} such as HeH, He$_2^+$, etc. By following Heitler and London, in 1930 Gentile evaluated the interaction energy between H and He, and between two helium atoms (including the polarization forces at the second order in the perturbation theory), and showed that normal He and H have no tendency to form a stable molecule \cite{Gentile1930}. Indeed, as recognized more in general one year later by Pauling \cite{Pauling1931}, resonance forces corresponding to the exchange of three electrons (with two electrons on a nucleus and one electron on the other) are mainly repulsive. This allowed him to formulate a rule for the occurrence of a three-electron bond:
\begin{quote} 
A three-electron bond, involving one eigenfunction for each of two atoms and three electrons, can be formed in case the two configurations A$: \!\! \cdot$B and A$\cdot \!\! :$B correspond to essentially the same energy \cite{Pauling1931}.
\end{quote}
In other words, both for the one-electron and for the three-electron cases, Pauling envisaged an exchange mechanism that was a direct generalization of what originally proposed by Heitler and London for molecules with identical atoms, where the resonance phenomenon involved degenerate (or nearly degenerate) electronic states. For the three-electron bond -- Pauling suggested -- this was the case for He$_2^+$, NO, NO$_2$, O$_2$, etc.

In the same 1931 paper, he also considered more specifically, but only {\it qualitatively}, the simple molecular ion He$_2^+$, by assuming, following Weizel \cite{Weizel1929} and Hund \cite{Hund1930}, that its formation was due to the bonding of a neutral helium atom with a ionized one, He + He$^+$ $\leftrightarrow$  He$_2^+$, in full analogy with the case of the hydrogen molecular ion H$_2^+$ (but with three electrons instead of only one), rather than with that of the compound HeH.

Actually, it may seem a contradiction that Pauling, convinced of the three-electron nature of the molecular bond in He$_2^+$ (``the simplest example of a molecule containing a three-electron bond is the helium molecule-ion, in which a $1s$ eigenfunction for each of two identical atoms is involved''), adopted a formation reaction similar to that of the one-electron bond molecule {\it par excellence}, the case of HeH -- though not realizing a stable bond for this specific compound -- being more suitable for his purposes. The reasoning of Pauling was to consider almost on the same ground the molecular bond in He$_2^+$ and He$_2$, this explaining the above association:
\begin{quote}
Evidence has been advanced \cite{Weizel1929} that the neutral helium molecule which gives rise to the helium bands is formed from one normal and one excited helium atom. Excitation of one atom leaves an unpaired $1s$ electron which can then interact with the pair of $1s$ electrons of the other atom to form a three-electron bond. The outer electron will not contribute very much to the bond forces, and will occupy any one of a large number of approximately hydrogen-like states, giving rise to a roughly hydrogen-like spectrum \cite{Pauling1931}.
\end{quote}
More clearly, by following Majorana \cite{MajoranaN2}, since the neutral helium molecule He$_2$ can be formed only from a normal (ground state) helium atom and an excited one, such a compound may undergo a dissociation into a neutral atom and a ionized one for sufficiently high energies of the excited electron, thus suggesting that the formation of the helium molecular ion occurs through the reaction above. 

Intriguing enough, Pauling did not explore the quantitative consequences of such an assumption in his 1931 paper, but we have to wait for about two years to see them \cite{Pauling1933}. This work was, however, carried out by Majorana in the same year 1931, about nine months before Pauling's first paper,\footnote{Note that Ref. \cite{MajoranaN2} was published in January, while Ref. \cite{Pauling1931} in September, 1931.} with a deeper mathematical formalism.

\section{The description of the helium molecular ion}

\noindent Majorana approached the problem of the possible formation of He$^+_2$ by starting from a discussion of the still unclear experimental result on the band structure in the helium emission spectrum, which brought spectroscopists to attribute the observed light bands just to the molecular ion He$^+_2$.

The basic physical idea was that of considering the system He$^+_2$ as similar to that of H$^+_2$, rather than HeH, i.e. the chemical reaction He + He$^+ \leftrightarrow$ He$^+_2$, but the fact that only one of the two atoms bonded to form the molecular ion is ionized made the problem at variance with what already known for the hydrogen molecule:
\begin{quote}
We want to study the reaction He + He$^+$ from the energy point of view and prove that such a reaction may lead to the formation of the molecular ion. [...] The method we will follow is the one that has been originally applied by Heitler and London \cite{HL} to the study of the hydrogen molecule. We shall assume that the electronic eigenfunctions of the molecule are linear combinations of the eigenfunctions belonging to the separate atoms and we shall use them to evaluate the average value of the interaction between the two atoms. However since the two nuclei have the same charge whereas only one of the atoms is ionized, the problem as we will show is mechanically rather different from the one discussed by Heitler and London and in general from the problem that one encounters in the normal theory of the homopolar valence \cite{MajoranaN2}.
\end{quote}
Majorana translated his idea in a suitable quantum mechanical theory by constructing appropriate eigenfunctions of the system in accordance with its symmetry properties, analogously to the group theory-inspired methods already adopted by him in other papers of his \cite{EspoAnn} \cite{EspoWeyl}.

Following Heitler and London, also Majorana started from the asymptotic solution of the problem (for large $R$) -- the wavefunctions of the system are just those for a neutral helium atom and its ion --, and as the former authors considered very unlikely that both electrons resided on the same nucleus in H$_2$ molecule, also the latter neglected the possibility that all three electrons in He$^+_2$ be located on the same nucleus. However, when the nuclei approach each other, their reciprocal interaction has to be taken into account, and such an interaction mixes all the wavefunctions previously introduced. Majorana was then able to recognize the only two appropriate combinations satisfying general symmetry principles, by emphasizing the relevance of {\it inversion} symmetry -- the total electronic wavefunction must show a definite symmetry with respect to the midpoint of the internuclear line:
\begin{quote}
To explain the chemical affinity between He and He$^+$ we must instead abandon the condition stated at the beginning and let the neutral atom free to share an electron with the ionized one and thus take its place. The net effect is to split the term resulting from the union of the two atoms, almost without changing its average value. The splitting thus depends not upon the resonance of the electrons but rather on the behavior of the eigenfunctions under reflection with respect to the center of the molecule. The eigenfunctions may be unchanged under the above spatial {\it inversion} in which case we call them {\it even}, or may change sign, in which case we call them {\it odd}. [...] The splitting of the term originating from its even or odd parity is greater by an order of magnitude than the energy due to the repulsive {\it valence} forces. One of the two modes of reaction thus corresponds to repulsion and does not give rise to chemical binding whereas the other gives rise to an attractive force and leads to the formation of a molecular ion. [...] We thus conclude that the essential cause of the instability of He$^+_2$ is the same which gives rise to
the stability of the molecular ion of hydrogen \cite{MajoranaN2}.
\end{quote}
Majorana then showed that two molecular states are possible for the He$^+_2$ molecular ion, only one of which corresponding to the bonding molecular orbital of the ion: such a configuration just reflects the fact that the ground state of the system is a resonance between the He$: \!\! \cdot$He and He$\cdot \!\! :$He configurations. More specifically, by denoting with $\Phi$ and $\varphi$ the unperturbed eigenfunctions of the neutral or ionized atom $a$, and similarly $\Psi$ and $\psi$ those of atom $b$, there are six eigenfunctions of the separate atoms obtained by permutations of the electrons and exchange of the nuclei:
\be
\ba{rclrclrcl}
A_1 \!\!\!\!&=&\!\!\!\! \varphi_1 \Psi_{23} , \quad  & A_2 \!\!\!\!&=&\!\!\!\! \varphi_2 \Psi_{31} , \quad  & A_3  \!\!\!\!&=&\!\!\!\! \varphi_3 \Psi_{12} , \\\ 
B_1 \!\!\!\!&=&\!\!\!\! \psi_1 \Phi_{23} , \quad       & B_2  \!\!\!\!&=&\!\!\!\! \psi_2 \Phi_{31} , \quad       & B_3  \!\!\!\!&=&\!\!\!\! \psi_3 \Phi_{12}.
\ea
\ee
However, the interaction among the two atoms mixes these states  ``according to the symmetry characters of the electrons and according to their behavior under spatial inversion'', so that we are left with only two singlet and two doublet states denoted by Majorana with $(\overline{123})^+$, $(\overline{123})^-$, $(\overline{12}3)^+$, $(\overline{12}3)^+$, respectively ($+$/$-$ labeling the even/odd terms), whose eigenfunctions are as follows:
\be
\label{hewf}
\ba{rcl}
y_1 \!\!&=&\!\! A_1 + A_2 + A_3 + B_1 + B_2 + B_3 , \\
y_2 \!\!&=&\!\! A_1 + A_2 + A_3 - B_1 - B_2 - B_3 ,\\
y_3 \!\!&=&\!\! A_1 - A_2 + B_1 - B_2 , \\
y_4 \!\!&=&\!\! A_1 - A_2 - B_1 + B_2 . 
\ea
\ee
The corresponding perturbed energy eigenvalues are, thus, symbolically wrote down in terms of several integrals involving the functions in (\ref{hewf}), and, just ``by taking into account the order of magnitude and the sign'' of such integrals, coming from symmetry and physical considerations and without explicitly evaluate them, Majorana concluded that the solution $y_3$ ``gives rise to repulsion whereas $y_4$ leads to the formation of a molecule.''

Now, in order to produce explicit numerical predictions for the potential energy curve, equilibrium distance, energy minimum and oscillation frequency of the helium molecular ion, Majorana made recourse to variational calculations, for which explicit expressions for the helium wavefunctions were required. 
\begin{quote}
The eigenfunction of the neutral atom of helium in its ground state has been calculated numerically with great accuracy but does not have a simple analytical expression. Therefore we need to use rather simplified unperturbed eigenfunctions \cite{MajoranaN2}.
\end{quote}
He wrote the ground state of the helium atom simply as the product of two hydrogenoid wavefunctions, but introduced an effective nuclear charge \cite{twoel} (as a variational parameter) describing the screening effect of the nuclear charge by means of the atomic electrons.\footnote{As explained in Ref. \cite{twoel}, what reported in the published paper does not reflect faithfully the great amount of work performed by Majorana on the helium wavefunctions, always devoted to get possible generalizations of the simple approximation just mentioned, giving easy but physically meaningful expressions.} Within this simple approximation, Majorana obtained a good agreement with the experimental data on the equilibrium internuclear distance and a not at all ``accidental'' (according to Majorana; see below) perfect agreement with the experimentally determined value of the vibrational frequency of He$^+_2$. Also, he estimated the dissociation energy of the molecular ion, obtaining a value $E_{\rm min}= - 2.4$ eV (including polarization forces effects) which Majorana could not compare with experiments, due to no available data, but which is {\it now} remarkably closer to the recent experimental determination of $E_{\rm min}= - 2.4457 \pm 0.0002$ eV \cite{Coman} than the recent theoretical prediction of $E_{\rm min}= - 2.47$ eV \cite{Ackermann}, obviously obtained with more refined mathematics than that used by Majorana.

As recalled above, similar quantitative results will be obtained independently\footnote{It seems that Pauling became aware of the Majorana paper around 1935 \cite{Pauling1935}, probably after the appearance of the important paper of  H.S.W. Massey and C.B.O. Mohr on transport phenomena in gases \cite{Massey1934}.} by Pauling two years later \cite{Pauling1933}, by adopting approximately the same variational procedure, with similar wavefunctions (again introducing an effective nuclear charge). However, it should be noted that Pauling did not employ a group theory-inspired method that allows to pick up the relevant terms in the wavefunctions, by exploiting the symmetry properties of the system, so that the underlying physical meaning was not as fully transparent as in Majorana. Also, a lucid analysis concerning the limits of applicability of the Heitler-London method to the present molecular system is as well present in Ref. \cite{MajoranaN2}, again clarifying the underlying physical meaning, which is worth to mention:
\begin{quote}
Heitler and London's method is inaccurate when the atoms are very far apart not only because it neglects the polarization forces, that in our case predominate, but also because it leads to resonance forces that are wrong both in the order of magnitude and in the sign \cite{MajoranaN2}.
\end{quote}
Finally, a simple comparison of the numerical results obtained by Majorana (discussed above) and by Pauling, concerning the equilibrium internuclear distance $r_0$, dissociation energy $E_{\rm min}$ and vibrational frequency $\omega_{1/2}$, with the experimental observations of the time (as reported in \cite{Pauling1933}),
\[
\ba{lclll}
{\rm Majorana:} & \qquad & r_0 = 1.16 \, {\rm A}, & E_{\rm min} = 2.4 \, {\rm eV}, & \omega_{1/2} = 1610 \, {\rm cm}^{-1} , \\{\rm Pauling:}    & \qquad & r_0 = 1.085 \, {\rm A}, & E_{\rm min} = 2.47 \, {\rm eV}, & \omega_{1/2} = 1950 \, {\rm cm}^{-1} , \\
{\rm experiments:}    & \qquad & r_0 = 1.090 \, {\rm A}, & E_{\rm min} = 2.5 \, {\rm eV}, & \omega_{1/2} = 1628 \, {\rm cm}^{-1} , 
\ea
\]
is rather amusing for the corresponding conclusions drawn by the two authors:
\begin{quote}
\begin{itemize}
\item[Majorana:] ``The calculation [...] is, quite by chance, in perfect agreement with the experimentally determined value'' \cite{MajoranaN2}.
\item[Pauling:] ``The experimental values [...] are in excellent agreement with the theoretical values'' \cite{Pauling1933}.
\end{itemize}
\end{quote}
Further similar calculations, performed with -- again -- different wavefunctions, introduced in order to have ``at infinity a correct wavefunction for He$^+$ and a reasonably good approximation (the screening-constant type) for He'' \cite{Weinbaum}, were later carried out by S. Weinbaum (and many others in the subsequent years), but without a remarkable improvement in the results obtained: the affair hopefully became just a numerical matter to be improved further and further.

\section{Ionic structures in homopolar molecules}

\noindent The successful description of the chemical bond of homopolar molecules, along the lines discussed above, did not close at all the corresponding chapter of quantum chemistry, since further experiments produced apparently conflicting results claiming for a thorough explanation. This was just the case when certain excited states of the hydrogen molecule were considered. 

Although a number of authors studied such states (see the review in Ref. \cite{Pauling1935}), here the unexplained phenomenon observed in the spectrum of the H$_2$ molecule was the decay of the excited $(2p \sigma)^2 \, ^1\Sigma_g$ ({\it gerade}) state into the ({\it ungerade}) $1 s \sigma \, 2p \sigma \, ^1\Sigma_u$ state in the infrared spectral region \cite{Weizel30}, contrarily to what happened in atomic systems, where the frequency corresponding to the transition $2p 2p - 1s 2p$ involving two excited optical electrons was very close to that of the transition $1s 2p - 1s 2s$ involving only one excited optical electron. 
\begin{quote}
Anomalous terms with both the electrons excited are known since a long time to occur in atoms with two valence electrons. In particular, the following are well known in numerous neutral or ionized atoms: $2p2p \, {}^3\!P_{012}$, $2p2p \, {}^1\!D$, $2p2p \,{}^1\!S$. According to a recent interpretation \cite{Weizel30} the $X$ term of the hydrogen molecule is formally analogous to these terms and should be precisely assigned to the configuration $(2p \sigma)^2 \, ^1\Sigma_g$. The analogy, however, breaks down in regard to the energies: whereas in atoms the frequency of the line $2p2p \rightarrow1s2p$ is of the same order of magnitude as the frequency of the line $1s2p \rightarrow1s2s$, the $X$ term is instead relatively deep, slightly above the normal term $(1s \sigma)(2p \sigma) \, ^1\Sigma_u$ with which it intensely combines in the infrared region; but the second one is in turn much higher than the ground state $(1s \sigma)^2 \, ^1\Sigma_g$ (ca. 12 volts) \cite{MajoranaN4}.
\end{quote}
The theoretical justification of even the existence of the $(2p \sigma)^2 \, ^1\Sigma_g$ term, along with an explanation of its abnormal energy level, when compared to similar atomic systems, then urged a reconsideration of the Heitler-London paradigm.

The problem was attacked early in 1931 (or, rather, at the end of 1930) by Majorana, who, after a lucid analysis of the known situation (see the quotation above), promptly recognized the relevant difference between atomic and molecular systems:
\begin{quote}
To consider such a state as a state with two excited electrons has purely formal meaning. In reality, to designate such terms with the states of the single electrons, though it may be convenient for their numbering and for the identification of those symmetry characters that are not affected by the interaction, does not allow by itself to draw reliable conclusions on the explicit form of the eigenfunctions. The situation is very different from the one of central fields [in atomic systems] where it is generally possible to neglect the interdependence of the electron motions (polarization) without loosing sight of the essentials \cite{MajoranaN4}.
\end{quote}
The misunderstanding was thus favored by the illegitimate transposition of atomic results into the the molecular framework, where neglecting the interaction between the electrons does alter the essential aspects of the phenomenon under study. 

Then, in order to explain the puzzling experimental results, Majorana  generalized the Heitler-London theory of the hydrogen molecule, where only configurations corresponding to one electron in each atom of the molecule were considered, by including different configurations where both electrons or no electron belong to a given atom. In other words, while Heitler and London considered only the chemical reaction H + H $ \leftrightarrow$ H$_2$ for the formation of the hydrogen molecule, Majorana introduced {\it also} the reaction H$^+$ + H$^-$  $\leftrightarrow$ H$_2$, where ionic structures are present. Of course, he was well aware of the fact that the apparent charge transfer via ionic structures has no proper physical interpretation in homopolar molecules and, for this reason, he designated such a reaction between the two hydrogen atoms as ``pseudopolar'' rather than ionic (the same term we used above -- ionic structure -- is borrowed by more modern valence bond approaches). Nevertheless he found that, while the normal $(1 s \sigma)^2 \, ^1\Sigma_g$ state predominately refers to the Heitler-London H + H reaction, 
\begin{quote}
the term $(2p \sigma)^2 \, ^1\Sigma_g$ [...] can be thought of as partially formed by the union H$^+$ + H$^-$. This does not mean, however, that it is a polar compound since, because of the equality of the constituents, the electric moment changes sign with a high frequency (exchange frequency) and therefore cannot be observed. It is in this sense that we speak of a pseudopolar compound \cite{MajoranaN4}.
\end{quote}
More in detail, by limiting himself to {\it gerade} singlet states (which is the case of both the $X$ terms and the ground state of the H$_2$ molecule), Majorana divided the configuration space into four regions -- $aa$, $ab$, $ba$ and $bb$ -- according to whether one or both electrons (labelled by $1,2$) are close to the nucleus $a$ or $b$. When the interaction between the two electrons is neglected, the four mentioned possibilities are equally represented in the given state, but, according to Majorana, ``the interaction increases the probability to find the system in $aa$ and $bb$, whereas it decreases that of $ab$ and $ba$'' \cite{MajoranaN4}. Indeed, qualitatively, the ground state of the molecule, dealt with by the Heitler-London theory, is described by a wavefunction whose major contribution comes from configurations $ab$ and $ba$, but, in order to explain the observations, a state must exist that is orthogonal to the ground state, whose main contribution then comes from configurations $aa$ and $bb$.

The chemical reaction between neutral atoms, H + H  $\leftrightarrow$ H$_2$, considered by Heitler and London, corresponds to configurations $ab$ and $ba$, described by an unperturbed wavefunction of the type
\be
y_2 = \varphi_1 \psi_2 + \varphi_2 \psi_1 ,
\ee
while configurations $aa$ and $bb$ originate from the reaction H$^+$ + H$^-$  $\leftrightarrow$ H$_2$ involving hydrogen ions, and are described by a wavefunction (symmetrized with respect to the exchange of the two nuclei) of the form
\be
y_1 = \Phi_{12} + \Psi_{12} .
\ee
Majorana then noted that the two states $y_1$ and $y_2$ are not orthogonal, but the ground state and the $X$ term of the H$_2$ molecule should result, in a first approximation, from two orthogonal combinations of them. The characteristic equation for determining the energy eigenvalues immediately followed, once an (approximate) expression for the wavefunction describing the H$^-$ ion, $\Phi_{12}$, was introduced semi-empirically\footnote{Majorana adopted the approximation employed by Hylleraas \cite{Hylle30} in his theoretical calculations for the solid lithium hydride.} to evaluate the intervening energy integrals. 

The numerical results for the equilibrium internuclear distance of the molecule in the $X$ state and for the corresponding vibrational frequency, that Majorana obtained from his impressive calculations, were amazingly close to the experimental observations, although he concluded his paper with his usual aloof and humble tone:
\begin{quote}
This result is even too favourable as, with the method we followed, we could have expected a value considerably smaller than the true one. [...] A quantitative evaluation is difficult but it is plausible that such an approximation tends to produce errors compatible with the discrepancies ascertained between calculation and experiment.
\end{quote}
Nevertheless, Majorana succeeded to prove the existence of a stable molecular state with both electrons in excited $2p$ orbitals, by improving the Heitler-London method to include ``pseudopolar'' interactions, thus paving the way for subsequent generalizations aimed at describing more complex molecular compounds.

\section{Conclusions}

\noindent The story of the quantum explanation of the nature of the chemical (homopolar) bond in molecules has been rather intricate, when compared to that of the quantum description of atoms. As a matter of fact, a number of authors contributed significantly to small or large pieces of such a story, by starting with the classification and ordering of spectroscopic data, just as in the way followed in atomic physics. However, successful theoretical methods developed in order to describe molecules revealed to be at variance with those employed for atoms, mainly due to the characteristic non-central problems to be treated. Methods of approximation able to find reliable solutions to the Schr\"odinger equation of the relevant molecular problem were developed, the Born-Oppenheimer method being the most important one, but novel ideas had to be introduced in order to gain more physical insight into the problem. As reviewed in Section 2, the Heitler-London approach (1927), complemented by the concept of molecular orbitals introduced by Lennard-Jones, was just the crucial starting point required for such a step, allowing the thorough description of the most simple molecules, H$_2^+$ and H$_2$, in terms of exchange interactions, along with the prediction of the non-existence of a stable helium molecule.

The (qualitative) description of different, more complex molecules required, then, the formulation of the concepts of one-electron and three-electron bonds introduced by Pauling (and described above in Sect. 3), with the {\it resonance} phenomenon considered by Heitler and London generalized to molecules with a number of valence electrons different from two. Apart from H$_2^+$, the first cases to be studied were those of the HeH compound and, especially, of the helium molecular ion He$_2^+$, for which Pauling himself envisaged a formation reaction of the type He + He$^+$ $\leftrightarrow$ He$_2^+$. The full quantitative description of the helium molecular ion was given in 1931 by Majorana (and, two years later, independently by Pauling), who extracted the appropriate symmetry properties of the molecular system considered and built a suitable quantum mechanical theory whose predictions revealed its astonishing success, as explained in Sect. 4.

It was due to Majorana also the apparently odd introduction of ionic structures into homopolar molecular compound, in order to explain puzzling experimental results regarding excited states of the hydrogen molecule. He thus generalized the Heitler-London method by including {\it pseudopolar} interactions able to account for the conflicting evidences as well as to improve phenomenological predictions for the given molecules. Seemingly, although a number of studies appeared on the ionic character of Hartree-Fock wavefunctions starting from 1949 \cite{Coulson} \cite{Clementi} \cite{Lowdin}, only in very recent times it has been recognized that ionic structures in the homopolar molecules yield binding energies predictions close to the experimental values \cite{CleCo}, while referring to {\it Majorana structures} as to ions that are not in the lowest ionic configuration \cite{Corongiu}.

While several other key contributions have appeared until now in the literature to improve out physical knowledge of the chemical bond, what discussed here remains the basic pillars upon which those improvements have been obtained and further results will come out in the near future.



\end{document}